\documentstyle[12pt,aaspp4,epsf]{article}

\begin{document}
\title{The Formation of Massive Stars by Accretion through Trapped Hypercompact
HII Regions}
\author{Eric Keto}
\affil{Harvard-Smithsonian Center for Astrophysics, 60 Garden Street, Cambridge
MA 02138}

\def\etal {{\sl et~al.\/}}
\begin{abstract}

The formation of massive stars may take place at relatively low accretion
rates over a long period of time if the accretion can continue
past the onset of core hydrogen ignition. The accretion may continue
despite the formation of an ionized HII region around the 
star if the HII region is small enough that the gravitational 
attraction of the star dominates the thermal pressure of the HII
region. The accretion may continue despite radiation pressure 
acting against dust
grains in the molecular gas if the momentum of the accretion flow is sufficient to
push the dust grains through a narrow zone of high dust opacity at the 
ionization boundary and into the HII region where the dust is sublimated.
This model of massive star formation by continuing accretion predicts
a new class of gravitationally-trapped,
long-lived hypercompact HII regions.  
The observational characteristics
of the trapped hypercompact HII regions can be predicted for 
comparison with observations.

This astro-ph version of the paper includes some corrections to the
equations in the ApJ version.

\end{abstract}


\section{Introduction}

The radiation of 
massive stars profoundly alters their immediate environment by ionizing 
the surrounding interstellar gas and exerting direct pressure on dust grains
in the gas.
In the formation of a massive star, there is only
a short time, on the order of $10^4$ yrs,
before a contracting massive protostar begins core hydrogen burning,
and the stellar radiation becomes significant.
Thus for accretion rates less than
$10^{-2}$ M$_\odot$ yr$^{-1}$,  
the outward forces of the thermal
pressure of the ionized gas and the radiation pressure on
dust grains will be competitive with the gravitational attraction
of the star during the accretion phase. 
If the effects of thermal and radiation pressure are capable of reversing the 
accretion flow within $10^4$ yrs, then the
formation of massive stars by accretion appears problematic.

Following previously suggested hypotheses as to how accretion
may proceed despite the effects of significant radiation,
this paper explores the possibility and consequences of the
formation of massive stars by accretion. 
In the context of simple models with spherical geometry,
the possibility of accretion through the thermal
pressure of 
HII regions was discussed in Walmsley (1995) and 
Keto (2002a, 2002b, hereafter K1 and K2), 
and the possibilities
for accretion through radiation pressure were
discussed in Kahn (1974), Yorke (1984, 2001), and 
Wolfire and Cassinelli (1987).
The effects of massive accretion flows onto the protostars themselves have
been studied in a series of papers by
Beech and Mitalas (1994),
Bernasconi and Maeder (1996),
Meynet and Maeder (2000),  Norberg and Maeder (2000), and
Behrend and Maeder (2001). 
In this paper, the relationships between these hypotheses are explored
to search for a consistent model for massive star formation.

This paper also examines some possibilities for observational confirmation of 
massive star formation by accretion. In particular,
the hypothesis of continuing accretion through trapped HII regions
predicts a new class of stable, long-lived hypercompact HII regions with
specific observable properties.
Some approximations 
allow an easy and accurate calculation of their expected free-free
radio continuum flux.

\section{The Trapped Hypercompact HII Regions}
\subsection{Introduction}

A recent theoretical study of the evolution of 
HII regions around newly formed stars (K2)
has found that if the stellar gravitational potential 
is included in the equations that describe the classical model for the pressure driven 
expansion of HII regions (Spitzer 1978; Dyson and Williams 1980; Shu 1992), 
the gravitational force is dominant over the thermal pressure at small scales.
These calculations show that a newly formed HII region 
cannot expand hydrodynamically if its size is smaller than a critical radius
($ r < GM/2c^2$) approximately where the escape velocity from
the star equals the sound speed, $c$, of the ionized gas.
The HII region is then trapped by the gravitational
field of the star.  Because the gravitational force is larger than the pressure force,
the molecular accretion flow responsible for the formation of 
the star is not significantly impeded by the thermal pressure
of the trapped HII region.
Rather the molecular flow becomes ionized on passing through the
ionization front and continues on toward the star.  
As a result of the continuing accretion,
the star will either increase in mass and
temperature or more stars will form until the flux of
ionizing photons is high enough that the equilibrium boundary of
ionization is beyond the critical radius for
hydrodynamic expansion. At this point, the HII region begins
a rapid transition to 
the evolution described
by the classical model of pressure-driven expansion. 
The beginning of the expansion essentially ends 
the accretion flow through the HII region.

\subsection{Size and time scales of the trapped HII regions}

Approximate size and time scales can be derived for the long-lived
hypercompact HII regions using as a first approximation
a simple model for steady-state spherical accretion.  In this model,
a star of mass $M$ with a flux of ionizing photons $N_i$ is
at the center of a spherically symmetric, steady accretion
flow driven by the gravitational attraction of the star. The 
self-gravity of the gas is ignored. The stellar radiation
maintains an ionized HII region around the      
star in equilibrium with the recombination rate within the HII region
and the mass flux of neutral gas through the HII region
boundary.
Both the neutral and ionized zones are assumed to be at
constant although different temperatures, and the equation of state in both
zones is assumed to be isothermal with sound speeds $c_1$ and $c_2$. 
With these assumptions the accretion flow
is described by the Bernouilli equation (Bondi 1952; or equation 2 of K2).

The maximum size of a trapped HII region is set 
by the critical radius where the relative velocity
of the static ionization front and the inwardly accelerating molecular
accretion flow is equal to approximately
twice the sound speed of the ionized gas, $2c_2$. 
This critical point always occurs just inside the 
sonic point, $r_c = GM/2c_2^2$, of the ionized
flow, and therefore one can say approximately that 
the HII region is unable to expand 
unless the sound speed of the
ionized gas exceeds the escape velocity from the star.
The sonic point is then a first estimate of the maximum 
size of a trapped HII region for
a star of a given mass (table 1).
If there is more than one
star in the HII region, for example a binary, the maximum size of the
HII region will scale proportionally
with the total mass enclosed within the HII region.

Since a trapped HII region is unable to expand hydrodynamically,
it will grow only as the flux of ionizing photons from the star
or stars at its center increases and the equilibrium radius of ionization
is found at increasingly larger distances.
The evolutionary time scale for the HII region in the trapped phase
is then set by the rate at which the stars gain mass.
In the spherically symmetric model, the accretion rate is,
\newcount\eqnum\eqnum=1
$$\dot M = 4 \pi \lambda (GM)^2 c_1^{-3}\rho_\infty
\eqno(\the\eqnum)$$
\newcount\accretionrate\accretionrate=\the\eqnum\advance\eqnum by 1
where $\lambda=1.12$, $c_1$ is the sound speed in the molecular gas,
and $\rho_\infty$ is the mass density of the gas at infinity.
Although the solution for the accretion flow is found assuming 
steady state,  if the star gains mass, the accretion rate 
will change over time. However, as long as the mass
does not change more quickly than the accretion flow can adjust, the
steady state approximation will be satisfactory.
The time, $t$, for the star to grow from
an initial mass, $M_i$, to a final mass, $M_f$, is,
$$t = {{c_1^3}\over{4\pi\lambda G^2\rho_\infty}}
   \bigg({{1}\over{M_i}} - {{1}\over{M_f}}\bigg)
\eqno(\the\eqnum)$$
\newcount\timescale\timescale=\the\eqnum\advance\eqnum by 1
or 
$$ t({\rm Myr})= 5.8\times 10^{4} {{1}\over{n_\infty ({\rm cm}^{-3})}}
\bigg( {{1}\over{M_i({\rm M}_\odot)}} - {{1}\over{M_f({\rm M}_\odot)}}\bigg)$$
This simple analysis indicates that the maximum size scales of 
the trapped HII regions 
are on the order of a hundred AU for single stars or a few hundred
AU for binaries and multiplets, and the minimum evolutionary time scales 
are $10^5$ to $10^6$ yrs.

\subsection{Evolution of the trapped HII regions}

In the preceding section the size and time scales for the trapped HII 
regions assumed that the HII region was at the maximum size allowed in the
trapped phase. 
The actual size of a trapped HII region will depend on the mass of the
star, its  ionizing
flux,  and the density of the molecular gas. The location of the ionization
front is determined by the equation of
ionization equilibrium which must be solved self-consistently with
the density and
velocity profiles of the ionized accretion flow set by the energy equation
governing the flow (\S 3.1 of K2).
The equation of ionization equilibrium yields three possible outcomes
for any given 
set of model parameters.
If the ionizing flux is weak 
relative to the accretion rate
($N_i m_H/\dot M < 1$), 
then
there will be no HII region. This situation corresponds to a ``quenched''
HII region. Secondly, if the ionizing flux and the
recombination rate balance at some radius less than the critical radius, the
result is a trapped HII region.
Or lastly, if the ionizing flux is strong enough that the radius of
ionization equilibrium is beyond the critical point, then the HII region
will be rapidly expanding by thermal pressure.
As a massive star at the center
of an accretion flow continues to gain mass, 
the ionization equilibrium may pass through all the
outcomes above as the flux of ionizing photons increases along with the
temperature and mass of the star. Tables 2 through 4 show a few cases illustrating
the evolution of an HII region
around single and multiple stars.

The tables  show that
depending on the surrounding molecular
density, the HII region begins ``quenched'', moves into the stable,
long-lived trapped phase, and finally moves beyond the critical radius into
the rapid expansion phase. 
Because stellar growth ends once the HII region enters
the expansion phase,
the final stellar type is determined entirely by 
the density of the molecular gas. 
At higher molecular densities, the star will grow to larger mass because
more ionizing radiation is required to move the boundary of the HII region
beyond the critical point for expansion.

Table 2 illustrates
the evolution of HII regions around  single stars. 
In table 3,
the mass due to stars within the HII region has been doubled
as could be caused by a few accompanying later type stars that add
mass to the system but do not increase the ionizing flux.
(The separation of the  group is considered to be small enough that the gravitational
potential of the system on the scale of the trapped HII region
is still essentially spherical.)
With more mass, the critical point of the accretion flow
moves outward and a higher ionizing flux is required to
end the evolution. 
In table 4, both the stellar mass and the
ionizing flux have been doubled, as might be the case for example with
an equal mass binary. 
But because the ionizing flux of
a star scales as a high power of its mass,  
massive stars forming together will have a lower flux-to-mass ratio 
than a single star of equivalent mass.
With a lower flux-to-mass ratio,
a group of stars will be able to continue accreting mass longer
and will grow to earlier spectral types
before their surrounding trapped HII region begins the expansion phase.
The trends in the tables show that the conditions that favor
the formation of the most massive stars are high densities of molecular gas
and formation in binaries or multiplets.  

Some caution must be used when interpreting the numbers in the tables.
Because the recombination rate scales as the square of the density, and
because the density increases rapidly inward asymptotically approaching
an $r^{-3/2}$ limit, the 
densities and the recombination rate become very high near
the stellar radius.
However, it is possible in real clouds that other effects
could limit the increase in density at very small radii. If so, the molecular
densities in tables 2 to 4 could be higher.

\section {Evolution of the stars}

Can a massive star of several tens of solar masses with a main 
sequence lifetime of the order $10^6$ yrs accrete sufficient mass 
before evolving
off the main sequence 
if the accretion rate were as low as those in table 1 ($10^{-5}$ to 
$10^{-4}$ M$_\odot$ yr$^{-1}$)?
To estimate the effect of the continuing accretion on the
stars powering the trapped HII regions, Alessandro Chieffi
(personal communication) ran numerical simulations of the
pre-main sequence and main sequence evolution of stars growing
to early types with continuous input of hydrogen at the stellar
surface (see Chieffi, Straniero, and Salaris (1995), and
Limongi, Straniero, and Chieffi (2000) for the method of
simulation). Figure 1 shows three evolutionary tracks of a 1 M$_\odot$ star
evolving with an accretion rate proportional to the square of the 
stellar mass (equation \the\accretionrate) and values of 0.33, 1.0, and 3.0
times $10^{-5}$ M$_\odot$ yr$^{-1}$
at 13 M$_\odot$. Also plotted is the evolution of a 1 M$_\odot$
star evolving to the main sequence without accretion 
(D'Antona and Mazzitelli 1996). Both stars follow an
identical path for some time before the effects of 
accretion cause a deviation toward increased luminosity. 
The higher luminosity is the result of  the larger radius of the
pre-main sequence star that is in turn caused by the burning 
of fresh deuterium continuously 
supplied by the accretion.
The track of the accreting star remains above the main sequence until
it has accumulated around 2.5 M$_\odot$. Thereafter
it evolves directly up the ZAMS line with increasing luminosity and 
temperature equivalent to luminosities and temperatures of non-accreting
ZAMS stars of progressively higher mass
(Schaller \etal\ 1992). The simulation shows that the mass accretion
rates in table 1 supply fresh hydrogen at a sufficient rate to
keep the accreting star on the ZAMS line. In other words, the star
does not exhaust its supply of hydrogen and remains on the main
sequence as long as the accretion continues. Thus relatively
slow accretion is able to build early type stars of high mass.

This result is the same as found earlier by Beech and Mitalas (1994),
Bernasconi and Maeder (1996), Meynet and Maeder (2000),
Norberg and Maeder (2000), 
and Behrend and Maeder (2001) for higher accretion rates than considered in
this paper.
Two evolutionary tracks given by Norberg and Maeder (2000)
are plotted in figure 1 for comparison. Compared to stars evolving
with the lower accretion rates of table 1, these tracks deviate earlier from
the evolutionary track of a non-accreting 1 M$_\odot$ star 
and join the main sequence at higher masses.
These higher rates
would require higher densities in the surrounding
molecular gas than considered in the examples in tables 1 through 4.
To achieve the rate in Norberg and Maeder (2000) 
of $10^{-5}$ M$_\odot$ yr$^{-1}$ onto a 
one M$_\odot$ star would require a density at infinity of 
$7\times10^4$ cm$^{-3}$ which implies a density $5\times 10^5$ cm$^{-3}$
at the sonic point located at 0.42 pc,
assuming a sound speed of 0.46 kms$^{-1}$ (T = 25 K).
The implied high densities are certainly possible in high mass star forming
regions, but as illustrated in the tables, single stars in such dense
surroundings would have 
difficulty in ionizing their HII regions beyond the critical radius
required to initiate expansion of the HII region and halt the accretion flow. 
Thus in such dense cores one might expect the formation of binaries or small 
goups of high mass stars.

Although the evolutionary tracks begin at 1 M$_\odot$, massive stars
are not likely to grow from such a low mass by Bondi accretion because
the time scale would be too long.  More likely the collapse of a dense 
molecular
cloud core will result in a star of some higher mass, for example
10 M$_\odot$,
that could then evolve to an earlier spectral type
by continuing accretion on a timescale of
a million years or so.

\section{Radiation Pressure}

Pressure on dust grains that absorb the 
radiation from the star or stars in an HII region and the radiation from any
shock at the base of an accretion flow where the flow decelerates can be 
competitive with the gravitational attraction of the stars
(Mestel 1954; Larson \& Starrfield 1971; Appenzeller \& Tscharnuter 1974;
Kahn 1974; Yorke and Krugel 1977; Wolfire and Cassinelli 1987)
One can derive a limiting luminosity-to-mass ratio by
balancing  the outward force of the radiation against
the inward force of gravity,
$${{L}\over{M}} = {{4\pi c G }\over {\kappa}}$$
Assuming an 
extinction coefficient for interstellar dust, $\kappa \sim 100 $ cm$^{2}$ g$^{-1}$
(Mathis, Rumpl, \& Nordsieck 1977),
the maximum luminosity to mass ratio is about 2500 L$_\odot$/M$_\odot$,
a value that would be exceeded by an early B type star. From this simple
argument, it would seem impossible to form a massive star by accretion.

However, this simple argument may not be correct. 
Kahn (1974) and Wolfire and Casinelli (1986, 1987)
have pointed out that the stellar radiation is absorbed in a thin
boundary layer that is found at the point in the accretion flow where the
gas temperature, that is increasing as the flow approaches the star,
reaches the sublimation temperature of the dust. In front of this layer,
the dust will be sublimated and therefore not available to absorb the
stellar radiation.  Behind this boundary layer,
the radiation field that is re-emitted by the dust will have the temperature of the
dust, not the star, and  the radiation will be mostly in the infrared.
Because the opacity of the dust at infrared wave lengths is
orders of magnitude lower than at optical and shorter wave lengths (Krugel \&
Siebenmorgan 1994) the transfer of momentum from the radiation to the dust will be
much lower. 
Thus for most of the cloud except within the absorbing boundary layer,
the infrared radiation passes through the cloud without exerting much pressure
on the dust.
At the boundary layer where the radiation temperature is the same
as the stellar photospheric temperature, 
the opacity of the dust is high and the dust will
absorb the full pressure of the radiation. However, because the boundary
layer is thin, it is possible for the momentum of the accretion flow to
push the gas and dust through this boundary layer into the sublimation
zone where the dust is destroyed. This requires that the momentum of
the flow $\dot M v$ must be higher than the momentum of the radiation given
by the ratio of the luminosity of the flow and the speed of light,
$L/C$.

The minimum momentum required to overcome the radiation pressure may be
estimated if we know the radius at which the dust sublimates and the 
velocity of the accretion flow at that radius. 
The sublimation
radius of the dust in a continuing accretion flow 
may be assumed to be the ionization boundary of the HII region.
The accretion flow speed is then given by the Bernouilli equation 
for the molecular gas. The minimum speed of the molecular flow at 
the boundary of a trapped HII region 
will be 
found when the HII region has reached its 
maximum trapped radius, approximately at the sonic point of
the ionized flow. At this radius, the molecular gas
has an infall velocity that is
always twice the sound speed of the ionized gas, and the minimum momentum of
the accretion flow onto an HII region in the trapped phase is the
flow rate times twice the sound speed of the ionized gas,
$\dot M 2c_2$.

Figure 2 compares the momentum of the radiation with the momentum of the flow for
different stellar masses. The minimum accretion rate for the HCHII
model is $\dot M > L/C2c_2$. 
The maximum rate in figure 2 is 
the Eddington luminosity, as a function of mass, 
where the absorption is due to 
classical electron scattering. 
The minimum and maximum accretion rates to overcome the radiation
pressure in the trapped HCHII model bracket the accretion rates
suggested in this paper as well as the higher rate suggested by Norberg and
Maeder (2000). The minimum accretion rate suggested by
Wolfire and Cassinelli (1987) is also plotted and appears
just below the minimum rate for the HCHII regions. The two estimates
are almost the same except that the sublimation radius 
of Wolfire and Cassinelli (1987), their equation 18, is calculated from more complex
considerations of the molecular gas temperature around massive stars
versus the sublimation temperature of grains; whereas, the sublimation
radius for continuing accretion is simply the ionization boundary of
the HII region.
The trend shown in figure 2 illustrates that
the accretion rate must increase as some power
of the stellar mass in the range of 1/2 to 2 in order
to overcome the radiation pressure and form stars of the highest
mass.

While the accretion flows considered in this paper are restricted to
spherical geometries to simplify the analysis,
other studies that consider radiation pressure in non-spherical flows 
suggest that the effect on the inflowing gas
is less than in spherical
geometries (Nakano 1989; Nakano, Hasegawa, Norman 1995; 
Jijina \& Adams 1996).

\section{Observational properties of the HCHII regions}

The trapped HCHII regions have a specific density structure, 
and it is possible
to predict their observational characteristics. Because they will be
found deeply
embedded in the dense molecular gas of massive star forming regions,
they will only be observable at long wave lengths from the infrared to radio.
Using some approximations it is possible to model the radio continuum
emission of the trapped HCHII regions
with very simple equations.

\subsection{Approximate densities and velocities}

The exact density profiles of the molecular and ionized gas in a
spherically symmetric accretion flow through a trapped HII region
are found by solving the non-linear Bernouilli equation for the energy of the 
accretion flow (equations 4 and 12 in K2). However, because the
accretion flows are dominated by gravity at radii less 
than the sonic point, $r_s=GM/2c^2$, and
by pressure at larger radii,
approximate densities can be obtained by using the free-fall density profile inside 
the sonic point and
the equation of hydrostatic equilibrium outside the sonic point.
If $n_{r_s}$ is the density at the sonic point, then at smaller radii
$$n(r) \approx n_{r_s} (r_s/r)^{1.5}
\eqno(\the\eqnum)$$
\newcount\powerlawapprox\powerlawapprox=\the\eqnum\advance\eqnum by 1
and at larger radii
$$n(r) \approx n_{r_s} exp \bigg[-2\bigg(1-{{r_s}\over{r}}\bigg)\bigg]
\eqno(\the\eqnum)$$
\newcount\hydrostaticapprox\hydrostaticapprox=\the\eqnum\advance\eqnum by 1
(The hydrostatic equation appears different from that in planetary
atmospheres because here the gravitational force varies with
radius as $1/r$ and is not assumed to be constant.)
Using these approximations the errors in density at
1/2, 1/10, 1/100 times the sonic radius are 33\%, 80\%, and 190\%.
On the outside at distances of 2, 3, and 4 time the sonic radius, the
errors are 32\%, 37\% and 39\%. Thus if the sonic point for a molecular
accretion flow is located at 0.1 pc, then over a range of radii from
0.001 pc to 0.4 pc the approximations are good to within a factor
of 2. The density at infinity that appears in
the scaling for the theoretical formulas is approximately $e^{-2}$ times 
the density at the sonic point.

A slightly better approximation to the density profile at smaller
radii can be obtained by
multiplying the approximate density by $\sqrt{2}$. This factor
partially compensates for the discrepancy between the approximate
and exact solutions near
the sonic point where the flow is still making a transition from hydrostatic to
free-fall. Just inside the sonic point,
the approximate power-law solution initially falls off too fast 
before become an excellent approximation, and thus 
the correct densities are slightly higher than
the approximate densities throughout the HII region.
The factor of $\sqrt{2}$ is 
arbitrary, but over the size scales appropriate
for the trapped HII regions, is an improvement.

Since the accretion flow is steady state, the velocity at any point 
in the accretion flow can be found from
the density profile using the equation of conservation and the fact that
by definition the velocity at the sonic point is the sound speed.
$$v(r) = {{r_s^2 n_{r_s} c}\over{ r^2 n(r)}}
\eqno(\the\eqnum)$$
\newcount\velapprox\velapprox=\the\eqnum\advance\eqnum by 1

These estimates suffice for the flows that are entirely within 
the ionized or molecular zones separately. For the complete flow
across the ionization boundary
it is possible to derive a simple approximate solution 
for the case where the boundary is at the sonic point of the
ionized flow. At the
maximum radius of an HII region in the trapped phase,
because the sonic point is so close to the
critical point, 
it is possible to ignore
the shock front 
that has
already begun to separate from the ionization front
and treat the HII boundary as a simple 
stationary ionization front with an R-critical jump (\S 3.2 K2).
In an R-critical jump, the velocity of the ionized gas leaving the front will
be equal to its sound speed. Knowing the velocity of both the ionized and 
molecular flows on either side of the boundary and knowing the 
density of the molecular gas,
it is possible to derive the density of the ionized gas from
the jump condition.
$$\rho_1 v_1 = \rho_2 v_2
\eqno(\the\eqnum)$$
\newcount\jumpcondition\jumpcondition=\the\eqnum\advance\eqnum by 1

Continuing the approximations, 
it is also possible to determine the ionizing flux in equilibrium
with the recombination rate within the HII region using the equation
for the radius of a Stromgren sphere (Spitzer 1978; or equation 13 of K2), 
$$ N_i = 4\pi \int^{r_s}_{r_*} n^2 \alpha r^2dr
\eqno(\the\eqnum)$$
\newcount\stromgren\stromgren=\the\eqnum\advance\eqnum by 1
but with a --3/2 power law density profile
rather than the constant density of the traditional Stromgren sphere. 
With $n\sim r^{-3/2}$,
$$ N_i = 4\pi r_s^3 n_{r_s}^2 \alpha [ ln(r_s) - ln(r_*)] 
\eqno(\the\eqnum)$$
\newcount\ionizingflux\ionizingflux=\the\eqnum\advance\eqnum by 1
Here $\alpha$ is the recombination rate, about $3\times 10^{13}$ cm$^{-3}$
s$^{-1}$, and $r_*$ is the stellar radius.
For the trapped HII regions
considered in this paper, the error in this approximation of the ionizing
flux is about 50\%.

As an example, consider 
a 40 M$_\odot$ star with molecular and ionized gas temperatures of 25 K (Wolfire
and Churchwell 1994)
and 10000 K. 
The sound speeds, $c_1$ and $c_2$ are then 0.46 kms$^{-1}$ and 12.9 kms$^{-1}$, and
the sonic points of the two flows are at $r_{s1}=0.42$ pc and $r_{s2} =$ 107 au.
Further assume a molecular gas density 
of 740 cm$^{-3}$ at $r_{s1}$.
(This implies a number density at infinity that is about
a factor of $e^{-2}$ less or about 100 cm$^{-3}$.)
Since we know the molecular velocities at $r_{s1}$ and $r_{s2}$ it is
better to calculate the number density using the conservation law
rather than the other way around.
The number density of the molecular gas at
$r_{s2}$ is 
$n(r_{s2}) = c_1 n_1 r^2_{s1} / 2 c_2 r^2_{s2} = 8.6\times 10^6$ cm$^{-3}$. 
Since the
velocity of the ionized gas at the back of the R-critical front at $r_{s2}$
is the sound speed, the jump condition,
$2n_1 v_1 = n_2 v_2$, yields the the density of the ionized gas at
$r_{s2}$, $n(r_{s2}) = 3.4\times 10^7$ cm$^{-3}$. 
(The factor of 2 in the jump condition assumes that the neutral gas
is composed of molecular hydrogen.)
The flux of ionizing photons required for photo-ionization equilibrium is then
$N_i = 3\times 10^{49}$, equivalent to the flux of an O5  star, assuming 
a stellar radius of 11.8 R$_\odot$. 

\section {
Radio emission of optically thick HII regions}

Observational studies have detected compact
sources of weak radio continuum emission
from the dense molecular
cloud cores that are thought to be the sites of massive star formation.
The compact radio continuum emission
is weaker than would be expected from optically thin ultracompact HII regions,
and its origin is not understood. Several possible explanations have been suggested:
dust in the HII regions may be absorbing ionizing photons; the ionizing
stars may not have reached the main sequence; the ionization may derive from
a cluster of less massive stars; the HII regions may be optically thick
(Wood \& Churchwell 1989b; 
Kurtz, Churchwell \& Wood 1994; Garay \etal\ 1993; Keto \etal\ 1994;
Miralles, Rodriguez \& Scalise 1994;
Garay \& Lizano 1999;
Carral \etal\ 1999, Molinari \etal\ 1998). Because  of their steep density
gradients, the HCHII regions will always be optically thick through their
central regions and their free-free emission could be the
observed weak continuum.
Using the  models of \S 2 for the trapped hypercompact HII regions, the 
characteristics of the free-free emission 
from the dense ionized gas in these sources can be predicted.
The radio free-free 
emission of a spherical HII region seen as an unresolved source is,
$$S_\nu = 4\pi k T_e \nu^2/c^2  
\int^{\theta_0}_0 \theta\big[1 - e^{-\tau_\nu}\big]d\theta
\eqno(\the\eqnum)$$
\newcount\exactemission\exactemission=\the\eqnum\advance\eqnum by 1
Here, $\theta$ is an angular coordinate on the plane of the
sky,
$$\theta = \sqrt{x^2 + y^2}/D$$ 
with coordinate axes $x$ and $y$ are centered on the HII region, 
$\theta_0 = R_0/D$ where
$R_0$ is
the radius of the HII region, and $D$ 
the  distance to the source.
An approximation to the free-free optical depth is given
by Altenhoff \etal\ (1960) or 
equation A.1a of
Mezger and Henderson (1967), 
$$\tau_\nu = 8.235\times 10^{-2} \bigg( {{T_e}\over{^\circ K}} \bigg) ^{-1.35}
	\bigg( {{\nu}\over{GHz}} \bigg)^{-2.1}
	\bigg( {{EM}\over{{\rm pc}\ {\rm cm}^{-6} }} \bigg)
\eqno(\the\eqnum)$$
\newcount\tauff\tauff=\the\eqnum\advance\eqnum by 1
where $T_e$ is the electron temperature, and $EM$ is the emission measure, $n_e^2L$.

Figures 3, 4 and 5  show the radio continuum emission as a
function of frequency for three HII regions of 107 au
with different densities.
For the gas density used in figure 3, the flux of ionizing photons
required for equilibrium is $10^{49.3}$ s$^{-1}$
and could be produced by a single 
06 star whose mass would be about 40 M$_\odot$ (Vacca, Garmany, and Shull 1996).
Since the recombination rate scales as the 
square of the density, the number of
ionizing photons for the models of figures 4 and 5 are factors of $10^2$ and 
$10^4$ less than the equilibrium flux of $10^{49.3}$ s$^{-1}$ of figure 3. 
The flux of figure 5 is quite low, but correspondingly,
the molecular densities consistent with the ionized density of the
model in figure 5 are much lower $<10^3$ cm$^{-3}$
than one would expect in a massive
star forming region.
Also plotted for comparison is the free-free radio emission that the HII
region would have if it had a constant density. 
The spectral indices of
the two cases show
that the trapped HII regions appear optically thick at lower frequencies 
because of the higher densities in their interiors. If one were to
derive the electron density from the continuum emission using assumptions of
constant density and optically thin emission, the constant density assumption
would overestimate the electron density while the thin assumption would
underestimate. A better procedure is outlined in the next section.

\subsection{Approximate fluxes for partially thick HII regions}

If a spherical HII region has a power law density gradient, then it will
always be optically thin along lines of sight with large impact
parameters that
have shorter pathlengths
and  pass through lower density gas. Conversely, it will always be
optically thick along lines of sight close enough to
the center.  
The flux from a such a partially optically thick HII
region can be estimated
by using the optically thin approximation on lines of sight at large
impact parameter where the optical depth is less than unity, and
the optically thick approximation on lines of sight near the center.
The approximate formulas 
provide an easy way to predict the
radio continuum emission from the trapped, hypercompact HII regions.

In the optically thin approximation, the brightness temperature is proportional
to the emission 
measure along the line of sight. In a spherical
HII region, a line of sight at impact parameter, $b$, will have
an emission measure,
$$EM(b) = 2\int^{Z_0}_0 n_e^2(r) dz
\eqno(\the\eqnum)$$
\newcount\EMone\EMone=\the\eqnum\advance\eqnum by 1
where 
$Z_0^2 = R_0^2 - b^2$, $R_0$ is the radius of the HII region, and
$2Z_0$ is then the path length or chord along the line of sight
at impact parameter, $b$. The Cartesian axes $x,y,z$ have their origin
at the center of the HII region, with $z$ along the line of sight, and
$x$ and $y$ in the plane of the sky.
If the electron density is $n(r) = n_0 (R_0/r)^{3/2}$ with $n_0$
being the density at the HII region boundary, the emission measure
is then,
$$EM(b) = 2R_0^3n_0^2 \int_0^{Z_0} (b^2 + z^2)^{-3/2} dz
\eqno(\the\eqnum)$$
\newcount\EMtwo\EMtwo=\the\eqnum\advance\eqnum by 1
which evaluates to,
$$EM(b) = 2 R_0^3 n_0^2 {{Z_0}\over{b^2(b^2 + Z_0^2)^{1/2}}}
\eqno(\the\eqnum)$$
\newcount\EMthree\EMthree=\the\eqnum\advance\eqnum by 1
Switching to angular coordinates, $\theta$, where D is the distance
to the source,
$$\theta = b/D$$
$$\theta_0 = R_0/D$$
$$EM(\theta) = 2Dn_0^2 {{\theta_0^2}\over{\theta^2}} \sqrt{\theta_0^2 - \theta^2}
\eqno(\the\eqnum)$$
\newcount\EMfour\EMfour=\the\eqnum\advance\eqnum by 1
For optically thin emission the brightness along the line of sight is 
directly proportional to the emission measure (equation 6 
of Mezger \& Henderson (1967),
$$	T_b(\theta) = A_{FF}\ EM(\theta)
\eqno(\the\eqnum)$$
\newcount\Tbright\Tbright=\the\eqnum\advance\eqnum by 1
The factor $A_{FF}$, 
assuming that the emission measure is in units of pc cm$^{-6}$, is,
$$ A_{FF} = 8.235  \times 10^{-2}
        T_e^{-0.35}
        \bigg( {{\nu}\over{GHz}} \bigg)^{-2.1}
\eqno(\the\eqnum)$$
\newcount\Aff\Aff=\the\eqnum\advance\eqnum by 1
For optically thick emission, the brightness temperature is simply equal to the
gas temperature.

The antenna temperature of an ideal antenna (assuming uniform illumination 
and ignoring inefficiency), is 
$$T_A = {{1}\over{\Omega}} 2 \pi \int ^{\theta_0}_0 \theta T_b(\theta) d\theta
\eqno(\the\eqnum)$$
\newcount\Tant\Tant=\the\eqnum\advance\eqnum by 1
where $\Omega$ is the solid angle subtended by the source.
If $\theta_1$ is the impact parameter at which the optical depth 
along a line of sight is unity, 
then using the equation for
optically thin emission for
the contribution to the antenna temperature from the part of 
the HII region exterior to $\theta_1$ 
$$T_A^{thin} = A_{FF}\ {{1}\over{\Omega}} 2 \pi \int_{\theta_1}^{\theta_0} 
           2Dn_0^2 {{\theta_0^2}\over{\theta}} \sqrt{\theta_0^2 - \theta^2}
\eqno(\the\eqnum)$$
\newcount\TAthinone\TAthinone=\the\eqnum\advance\eqnum by 1
which evaluates to,
$$T_A^{thin} = -4 A_{FF}\ D n_0^2 \bigg(u + {{\theta_0}\over{2}} 
	ln{{\theta_0 - u}\over{\theta_0 + u}} \bigg) 
\eqno(\the\eqnum)$$
\newcount\TAthintwo\TAthintwo=\the\eqnum\advance\eqnum by 1
where,
$$u = \sqrt{\theta_0^2 - \theta_1^2} $$
The contribution to the antenna temperature from 
the thick part of the disk, $\theta < \theta_1$, using the equation for
optically thick emission is,
$$T_A^{thick} = {{1}\over{\Omega}} 2 \pi \int ^{\theta_1}_0 T_e d\theta
\eqno(\the\eqnum)$$
\newcount\TAthickone\TAthickone=\the\eqnum\advance\eqnum by 1
which evaluates to,
$$T_A^{thick} = {{\theta_1^2}\over{\theta_0^2}} T_e
\eqno(\the\eqnum)$$
\newcount\TAthicktwo\TAthicktwo=\the\eqnum\advance\eqnum by 1
The antenna temperature for the whole  HII region may be approximated as,
$$T_A = T_A^{thin} + T_A^{thick}
\eqno(\the\eqnum)$$
\newcount\TAtotal\TAtotal=\the\eqnum\advance\eqnum by 1
or in terms of flux density,
$$S_\nu = 2kT_A\Omega \nu^2/c^2
\eqno(\the\eqnum)$$
\newcount\fluxdensity\fluxdensity=\the\eqnum\advance\eqnum by 1

The value of $\theta_1$ can be determined from the equation for
the optical depth of the free-free emission,
$$\tau_\nu(\theta) = 
    A_{FF}/T_e 2Dn_0^2{{\theta_0^2}\over{\theta^2}}\sqrt{\theta_0^2 - \theta^2}
\eqno(\the\eqnum)$$
\newcount\taufftwo\taufftwo=\the\eqnum\advance\eqnum by 1
With $\tau_\nu = 1$, the solution is,
$$\theta_1^2 = 2(A_{FF}Dn_0^2\theta_0^2/T_e)^2\bigg(-1 
     + \sqrt{1 + \theta_0^2(A_{FF}Dn_0^2\theta_0^2/T_e)^{-2}}\bigg)
\eqno(\the\eqnum)$$
\newcount\taufthree\taufthree=\the\eqnum\advance\eqnum by 1
Table 5 lists fluxes 
determined both by this approximate method with the approximate density
profile of \S 2.1, and by
numerical calculation 
using the exact density profile as given
by the Bernouilli equation. The example HII region in table 5 has the same
structure as used to compute the fluxes
in figure 4. Table 5 shows that the approximation is correct to
within 35\% whereas the optically thin approximation, also shown in the
table, is off by orders
of magnitude when the emission is partially optically thick.
Thus derivations of the electron density and the number of ionizing photons
in small HII regions with steep density gradients
may be in error if the emission measure is determined 
using the optically thin approximation over the entire HII region.

\section{Broad radio recombination lines}

Radio recombination lines
observed in hypercompact HII regions  tend to
be broader  (50 to 180 kms$^{-1}$) than those observed
in typical UCHII regions  (30 kms$^{-1}$) 
(Altenhoff, Strittmatter \& Wendker 1981; Zijlstra \etal\ 1990;
Keto \etal\ 1995; 
Gaume, Fey \& Claussen 1994; De Pree \etal\ 1994, 1996, 1997;
Jaffe \& Martin-Pintado 1999; Keto 2001;). 
The large line widths have generally been interpreted 
as evidence for high velocity outflows. This is certainly the
case for some HII regions, but the model for continuing
accretion offers another interpretation
applicable to those HII regions that are in the trapped phase.
In the trapped hypercompact HII regions, the large recombination line 
widths are due to a combination of
pressure broadening and high infall velocities in the steep density
gradients and accelerating accretion flows through the trapped HII regions.

The lowest infall velocity in the ionized gas of a trapped HII region 
is the sound speed and the ionized accretion flow will attain this 
velocity
at the ionization boundary when the HII region is at its
maximum trapped size. At smaller radii and for smaller HII regions the
velocity scales inward as approximately $r^{-1/2}$. Thus
infall velocities of many times the sound speed are realizable.
The ratio of pressure broadening to thermal broadening scales linearly 
with the electron density and as the seventh power of principal quantum
number of the recombination line. 
At cm wavelengths the pressure broadening is 
equivalent to the thermal linewidth at densities as 
low as $10^5$ cm$^{-3}$ (Keto \etal\ 1995),
$${{\Delta\nu_p}\over{\Delta\nu_{th}}} = 1.2 {{n_e}\over{10^5 ({\rm cm}^{-3})}} 
\bigg( {{N}\over{92}} \bigg)^7$$
As the density increases inward even more steeply
than the velocity, pressure broadened widths of many times the sound speed
are to be expected in hypercompact HII regions.

An interesting correlation
between the spectral index of the free-free radio continuum
and the line width of the H66$\alpha$ was noted in 
the hypercompact HII regions in the W49A star forming region
(De Pree \etal\ 1997). This correlation  
would be expected if the recombination line widths
had a component due to pressure broadening since both pressure 
broadening and
the spectral index are higher for high density gas. A lack of
correlation between linewidth and electron density is also expected
if the electron densities are calculated assuming constant density and
optically thin emission while the HII regions do not meet these
assumptions because of steep density gradients.

\section{The life times of UCHII region }

An HII region in the classical model for pressure driven evolution
expands at approximately the sound speed of the ionized gas, and
the age of an expanding HII region is approximately equal 
to its radius divided by the sound speed. Expanding ultracompact HII regions
observed to have radii less than 0.1  pc would then have dynamical 
ages of a few
thousand years. Such short ages are inconsistent with 
observations.  Because the gravitationally trapped HII regions have much
longer life times, they potentially resolve a
number of these observational problems. However, not all HII regions are 
gravitationally trapped, and this model will not be appropriate for
many observed HII regions. In particular, HII regions with 
cometary or arc-shaped
morphologies are likely to be rapid outflows, ``champagne'' flows, 
down steep density gradients
in the surrounding molecular gas (Tenorio-Tagle 1979;
Franco, Tenorio-Tagle \& Bodenheimer 1990;
Keto \etal\ 1995; Shu \etal\ 2002).
HII regions with irregular or shell-like 
morphologies are not well explained, but are unlikely candidates for trapping.

\subsubsection{The embedded population}

The number of HII regions in the galaxy has been estimated
by Wood and Churchwell (1989a,b) who identified embedded HII
regions by a combination of characteristic 
infrared colors and fluxes in the IRAS data base 
(infrared luminosities $>10^4$ L$_\odot$ and
colors $25{\rm m}\mu/12{\rm m}\mu > 3.7$
and $60{\rm m}\mu/12{\rm m}\mu > 19.3$). They found  1717 such sources,
and by comparison with the population of visible O stars, determined
that O stars spend 10 to 20\% of their lifetime deeply embedded in the
molecular clouds. This implies a lifetime for the embedded phase
of $10^5$ yrs or more, incompatible with the $10^3$ yr dynamical time scale
of a pressure driven HII region, but consistent with the expected time scale of
a trapped HCHII region that is slowly growing on 
an accretion time scale.
Many of the Wood and Churchwell IRAS sources may be embedded trapped
HCHII regions.

\subsubsection{Time scales for formation in clusters}

Continuum observations of massive star forming regions often show
clusters of ultracompact HII regions, many with size scales of
0.01 to 0.1 pc. The dynamical ages of these HII regions in 
pressure-driven expansion of $10^3$ to $10^4$ yrs are significantly
shorter than the gravitational free-fall time ($10^5$ yrs) for molecular gas
with a density typical of massive star forming regions 
of $10^5$ cm$^{-3}$. The short ages implied by the model
for pressure-driven expansion raise the question of how
the stars powering these HII could have formed simultaneously with
a spread in age that is only a fraction of the free-fall time 
of the host molecular core (Ho, Klein \& Haschick 1987). 
The gravitationally trapped HCHII regions resolve this problem
because their life times
are tied directly to the accretion time scale of the stars.

\subsubsection{Time scales for expansion in clusters}

The presence of multiple UCHII regions in clusters indicates 
that most of the individual HII
regions in a cluster cannot be rapidly expanding. Some of these
UCHII regions, 
despite their relatively large sizes,
may be gravitationally trapped 
by a small group of stars. For example,
an HII region that has
a size of 0.01 pc could be trapped by a total stellar mass of 800 M$_\odot$.
The HII region G10.6--0.4 (K1) is an example of an 
UCHII region gravitationally trapped by a small group of stars
whose combined mass totals several hundred M$_\odot$. 
In other cases, what appears to be 
an UCHII regions may really be an unresolved 
group of HCHII regions each of which is small enough to
be trapped by a single star or binary. These individual HCHII regions
only become distinct when viewed at sufficiently high angular resolution. 
Examples in the literature include the massive-star forming clusters 
Sgr B2 (Gaume \etal\ 1995)
and W49 (Depree \etal\ 1997). Figures 1, 2 and 4 in Gaume \etal\ (1995)
show that in Sgr B2,  a few UCHII regions 
with apparent diameters of 0.1 to 0.2 pc seen
in a centimeter continuum
image at an angular resolution of $2.5^{\prime\prime}$ 
pc are resolved into
about 50 HCHII regions with sizes less than 
5000 AU when observed at 10 times better resolution. 
Furthermore, the individual HII regions seen at the
higher angular resolution may themselves be
composed of still smaller HII regions.
The kpc distances of most
massive star forming regions makes it difficult to determine the true size scales
of the individual HII regions in a cluster.

\section{HII regions that are not trapped}

Hypercompact and ultracompact HII regions 
may in general be in a variety of dynamical
states. 
For example, tables 2 through 4 illustrate that the state of an
HII region depends sensitively on
the density of the surrounding molecular gas, but
this model is predicated on a particular and idealized density
profile that will not be found in real molecular cores.
If a massive star forms in a molecular core that
for whatever reason has a density gradient steeper than
$r^{-3/2}$, ionization equilibrium is not possible 
(Franco, Tenorio-Tagle \& Bodenheimer 1990), and a continuous-gradient version of the 
champagne outflow will result 
(Tenorio-Tagle 1979;Keto \etal\ 1995; Shu \etal\ 2002; Lizano \etal\ 2003). 
Thus not all hypercompact HII will go through a trapped phase.
Conversely the trapped phase is not necessarily 
restricted to hypercompact HII regions.
Trapped ultracompact HII 
regions may form around a tight cluster of early stars whose
combined mass is sufficient to establish a sonic point
relatively far from the cluster 
(K1). 

The morphology of an HII region can offer some guidance
to the dynamical state.
For example, HII regions that are cometary or shell in shape
are unlikely to be trapped. The cometary morphology is suggestive
of expansion down a steep density gradient while the shell
morphology implies a density structure that is the result 
of other outward forces such as stellar winds.
However, the dynamics within or outside a trapped HII region
need not be strictly symmetric
because the confinement due to the gravitational force
of the stars 
is inherently hydrodynamically stable. 
For example, because of this stability,
it should be possible for a trapped HII region to co-exist with other
phenomenon such as an accretion disk or bipolar outflow. 
Nevertheless, the most likely morphology of a trapped HII region
would be roughly spherical and centrally bright.

\section{Conclusions}

\noindent It is possible to form high mass stars by accretion despite
the thermal and radiation pressure created by the radiation of the
high mass stars and the radiation from the shock at the base of the
star forming accretion flow.

\noindent Accretion can continue past the time of formation of an
HII region around a newly formed star if the equilibrium radius 
of ionization is small enough that the escape velocity from the star
exceeds the sound speed of the ionized gas at the HII region
boundary.

\noindent Radiation pressure will not reverse a star forming accretion
flow if the momentum of the flow at the HII region boundary
exceeds the momentum of the radiation. In this case the dust grains are
pushed through the ionization boundary and sublimated.

\noindent ZAMS stars within the trapped HII regions may grow to higher masses
and earlier spectral types by continuing accretion through the HII region.
The evolutionary path of the accreting stars follows directly up the ZAMS line.

\noindent The final mass that a star attains within a trapped HII region will
depend on the density of the surrounding molecular gas.

\noindent The conditions that favor the formation of the most massive stars
are high densities in the surrounding molecular gas and formation 
with one or more partners in a binary or multiplet.

\noindent The radio free-free emission of the trapped hypercompact
HII regions will always be optically thick through the center
of the HII region. High optical depth and small size ensure that
the radio emission will be relatively weak.

\noindent The long lived gravitationally trapped phase 
addresses the
difficulties imposed by the model for the pressure driven
expansion of HII regions that predicts impossibly short
lifetimes for hypercompact and ultracompact HII regions.

\clearpage

\begin{deluxetable}{rrrrrrrrr}
\tablenum{1}
\tablecolumns{6}
\tablewidth{0pc}
\tablecaption{Parameters for Trapped HII Regions}
\tablehead{
\colhead{Spectral Type} & \colhead{Mass$^a$}   & \colhead{UV flux$^a$} 
		& \colhead{Maximum radius$^b$}  &\colhead{Accretion time$^c$} 
		& \colhead{Accretion Rate}\\ 
\colhead{} & \colhead{(M$_\odot$)}   & \colhead{(log s$^{-1})$} & \colhead{(au)}   
		& \colhead{Myr} & \colhead{ ($10^{-4}$ M$_\odot$ yr$^{-1}$)}\\ 
}
\startdata

B0.5	&	18.4	&	47.90	&	49.5	&	0.0	&	0.47	\\
B0	&	19.5	&	48.16	&	52.5	&	0.02	&	0.53	\\
09.5	&	20.8	&	48.38	&	56.0	&	0.05	&	0.60 	\\
09	&	22.1	&	48.56	&	59.4	&	0.06	&	0.68 	\\
O8.5	&	23.6	&	48.72	&	63.5	&	0.09	&	0.77 	\\
O8	&	25.1	&	48.87	&	67.5	&	0.10	&	0.88	\\
O7.5	&	26.9	&	49.00	&	72.4	&	0.12	&	1.01	\\
O7	&	28.8	&	49.12	&	77.5	&	0.14	&	1.15	\\
O6.5	&	30.8	&	49.23	&	82.9	&	0.16	&	1.32	\\
O6	&	33.1	&	49.34	&	89.0	&	0.17	&	1.52	\\
O5.5	&	35.5	&	49.43	&	95.5	&	0.19	&	1.75	\\
O5	&	38.1	&	49.53	&	102.5	&	0.20	&	2.01	\\

\enddata

\tablecomments{
\hfill\break $^a$ Vacca, Garmany \& Shull (1996)\hfill\break
$^b$ Assumes a temperature of $T=10^4$ K for the ionized gas
and 25 K for the molecular gas. \hfill\break
$^c$ The accretion times assume a number density at infinity of 1000 cm$^{-3}$
(equation \the\timescale).
}

\end{deluxetable}

\begin{deluxetable}{rrrrrrrrr}
\tablenum{2}
\tablecolumns{4}
\tablewidth{0pc}
\tablecaption{Evolution of HII regions around single stars}
\tablehead{
\colhead{Spectral Type} 
			& \colhead{$n_\infty =150 $ cm$^{-3}$}   
                        & \colhead{$n_\infty =200 $ cm$^{-3}$} 
			& \colhead{$n_\infty =250 $ cm$^{-3}$}  \\
\colhead{} 	& \colhead{(au)}   
		& \colhead{(au)} 
		& \colhead{(au)}   \\
}
\startdata

B0.5	&	1 	&	Q	&	Q   	\\
B0	&	3 	&	Q    	&	Q   	\\
09.5	&	10 	&	1    	&	Q   	\\
09	&	26  	&	3    	&	1   	\\
O8.5	&	46  	&	5    	&	1   	\\
O8	&	94  	&	10   	&	2   	\\
O7.5	&	E 	&	15 	&	2   	\\
O7	&	E   	&	21  	&	3   	\\
O6.5	&	E   	&	27   	&	4   	\\
O6	&	E   	&	32   	&	4   	\\
O5.5	&	E   	&	33   	&	4   	\\
O5	&	E   	&	36   	&	5    	\\

\enddata

\tablecomments{
\hfill\break
Evolution of an HII region around a single star that is accreting
mass and growing in spectral type. The numbers in the table
indicate the radius of the trapped HII region around the star.
The letter Q indicates that the HII region is ``quenched'' meaning that there
is no solution for ionization equilibrium because the ionizing
flux is too small compared to the accretion rate. The letter E means
that the HII region is no longer trapped, but expanding. The growth
of the star stops once the HII region begins expanding.
}

\end{deluxetable}

\begin{deluxetable}{rrrrrrrrr}
\tablenum{3}
\tablecolumns{4}
\tablewidth{0pc}
\tablecaption{Evolution of HII regions around multiple stars}
\tablehead{
\colhead{Spectral Type} & \colhead{$n_\infty = 50$ cm$^{-3}$}   
                        & \colhead{$n_\infty = 75$ cm$^{-3}$} 
			& \colhead{$n_\infty = 100$ cm$^{-3}$}  \\
\colhead{} 	& \colhead{(au)}   
		& \colhead{(au)} 
		& \colhead{(au)}   \\
}
\startdata

B0.5	&	1	&	Q	&	Q   	\\
B0	&	5  	&	Q   	&	Q   	\\
09.5	&	20  	&	1    	&	Q   	\\
09	&	57   	&	2   	&	Q   	\\
O8.5	&	101  	&	3   	&	Q   	\\
O8	&	E     	&	6    	&	1  	\\
O7.5	&	E    	&	10   	&	1  	\\
O7	&	E   	&	14  	&	2  	\\
O6.5	&	E   	&	18   	&	2  	\\
O6	&	E     	&	22   	&	2  	\\
O5.5	&	E   	&	22   	&	2  	\\
O5	&	E   	&	25   	&	2   	\\

\enddata

\tablecomments{
\hfill\break
Evolution of an HII region around a group of stars that is accreting
mass and growing in spectral type. The calculation assumes that there
are a number of later type stars with non-ionizing radiation accompanying
one massive star. In these examples, 
the combined mass of the later type stars is set
equal to the mass of the early type star. In other words, relative to
table 2, the mass has been doubled, but the ionizing flux has remained
the same. The numbers and letters have the same meaning as in table 2. 
Relative to table 2, the higher total mass allows the formation of 
individual stars of higher mass in lower densities of molecular gas.
}

\end{deluxetable}

\begin{deluxetable}{rrrrrrrrr}
\tablenum{4}
\tablecolumns{4}
\tablewidth{0pc}
\tablecaption{Evolution of HII regions around equal mass binary stars}
\tablehead{
\colhead{Spectral Type} & \colhead{$n_\infty = 50$ cm$^{-3}$}   
                        & \colhead{$n_\infty = 75$ cm$^{-3}$} 
			& \colhead{$n_\infty =100$ cm$^{-3}$}  \\
\colhead{} 	& \colhead{(au)}   
		& \colhead{(au)} 
		& \colhead{(au)}   \\
}
\startdata

B0.5	&	16	&	1	&	Q   	\\
B0	&	83	&	3    	&	Q   	\\
09.5	&	E  	&	11   	&	1   	\\
09	&	E   	&	37   	&	3   	\\
O8.5	&	E   	&	72    	&	5   	\\
O8	&	E   	&	115   	&	11   	\\
O7.5	&	E    	&	140  	&	17   	\\
O7	&	E	&	E   	&	25   	\\
O6.5	&	E   	&	E    	&	33   	\\
O6	&	E   	&	E    	&	41   	\\
O5.5	&	E   	&	E    	&	41  	\\
O5	&	E   	&	E    	&	46    	\\

\enddata

\tablecomments{
\hfill\break
Evolution of an HII region around a pair of stars of equal mass. 
Both stars are accreting
mass and growing in spectral type. 
Relative to
table 2, both the mass and the ionizing flux have been doubled.
The numbers and letters have the same meaning as in table 2. 
Relative to table 2, the higher mass and ionizing flux allows 
the formation of larger trapped HII regions.
}

\end{deluxetable}

\begin{deluxetable}{rrrrrrrrr}
\tablenum{5}
\tablecolumns{6}
\tablewidth{0pc}
\tablecaption{Radio continuum flux of the trapped hypercompact HII regions}
\tablehead{
\colhead{Frequency} & \colhead{$r(\tau=1)$ exact$^a$}   
                        & \colhead{$r(\tau=1)$ approx.$^b$} 
			& \colhead{Exact Flux$^a$}  
			& \colhead{Approx. Flux$^c$} 
			& \colhead{Thin Flux$^d$}  \\
\colhead{(GHz)} 	& \colhead{(au)}   
		& \colhead{(au)} 
		& \colhead{(mJy)} 
		& \colhead{(mJy)} 
		& \colhead{(mJy)} \\
}
\startdata

  1.4  &   107.0  $^e$&     107.0 &0.06&   0.06&   256 \\
  5.0  &       107.0  &     107.0 &0.7 &   0.7 &   226 \\
 15.0  &       105.9  &     103.5 &6   &   9   &   202 \\
 23.0  &       101.6  &     92.8  &13  &   19  &   194 \\
 44.0  &       71.7   &     60.5  &33  &   42  &   182 \\
100.0  &       26.7   &     27.7  &63  &   64  &   167 \\
230.0  &       9.6    &     11.7  &85  &   91  &   154 \\
345.0  &       5.9    &     7.7   &93  &   102 &   148 \\
690.0  &       2.7    &     3.7   &102 &   138 &   120 \\

\enddata

\tablecomments{
\hfill\break
$^a$Calculated numerically using density profile as given by the
solution of the Bernouilli equation.\hfill\break
$^b$Radius where $\tau = 1$ using approximate method of \S 5.1.\hfill\break
$^c$The approximate fluxes using the approximate model for 
trapped hypercompact HII region as described in \S 2.1, and the approximate
formula for the flux as described in \S 5.1. The ionized gas densities include
the $\sqrt{2}$ correction.
\hfill\break
$^d$Flux calculated using the optically thin approximation.\hfill\break
$^e$ At low frequencies the HII region is essentially optically 
thick everywhere.
}

\end{deluxetable}

\clearpage

\begin{figure}
\epsscale{0.6}
\plotone{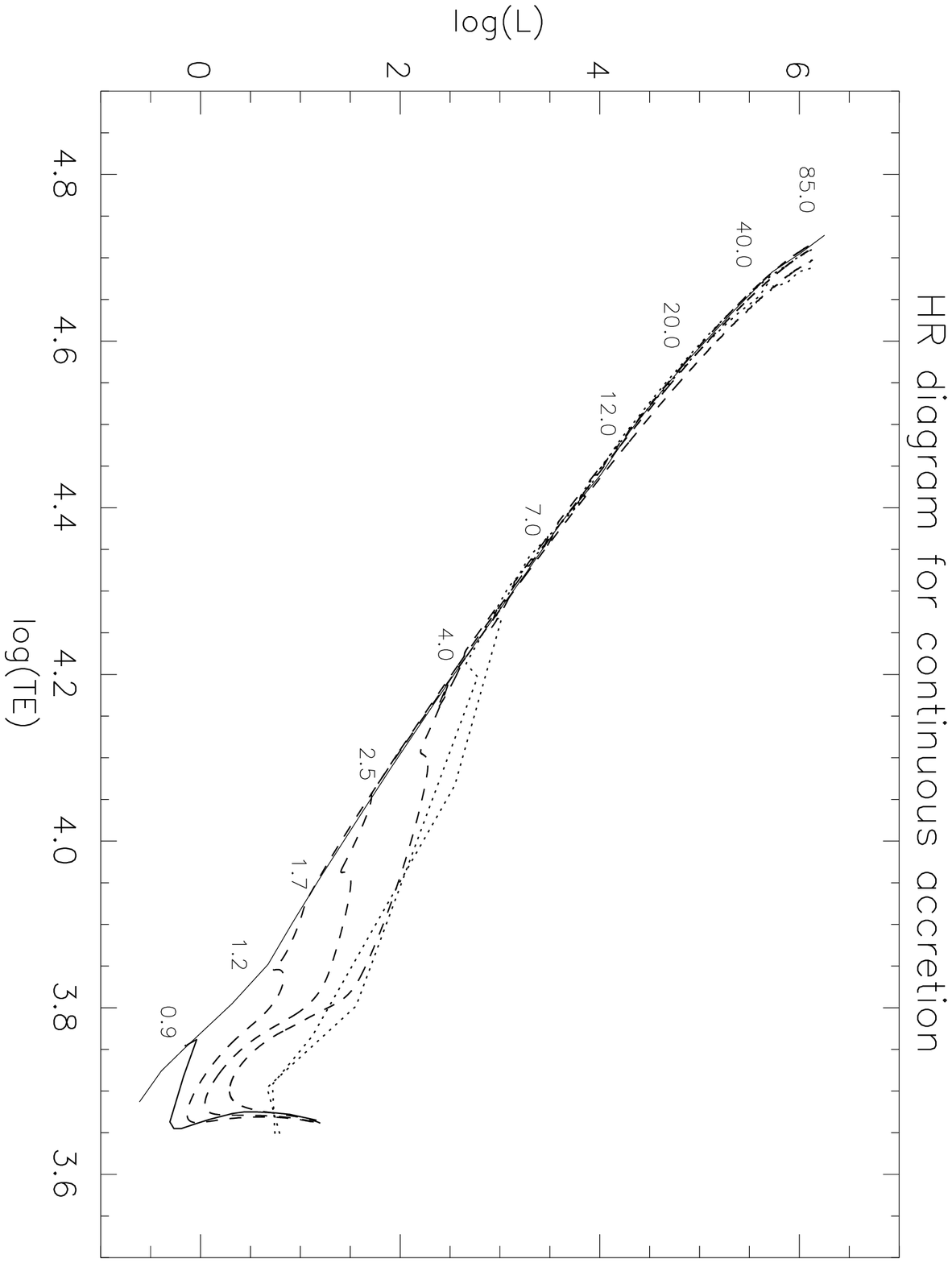}
\caption{
HR diagram for massive stars forming by accretion. The ZAMS line (solid line)
is from
Schaller \etal\ (1992). The three dashed lines are the track of a star
accreting mass at three different rates, each
proportional to the stellar mass squared as suggested
by the simple model of \S 2. The
accretion rates are set to  0.33, 1.0, and 3.0 times
$10^{-5}$ M$_\odot$ yr$^{-1}$ for a stellar mass of
13 M$_\odot$. The evolution starts at 1 M$_\odot$ and initially follows
the evolutionary track (solid line)
of a non-accreting star (D'Antona and Mazitelli 1994) before diverging to
higher luminosity.
The two thin dotted lines (from Norberg and Maeder 2000)
show the evolution of stars accreting mass at
a higher rate, $10^{-5}$ M$_\odot$ yr$^{-1}$ for a stellar mass of 1 M$_\odot$
and increasing as the 1.0 and 1.5 power of the stellar mass.
}

\end{figure}

\begin{figure}
\epsscale{0.6}
\plotone{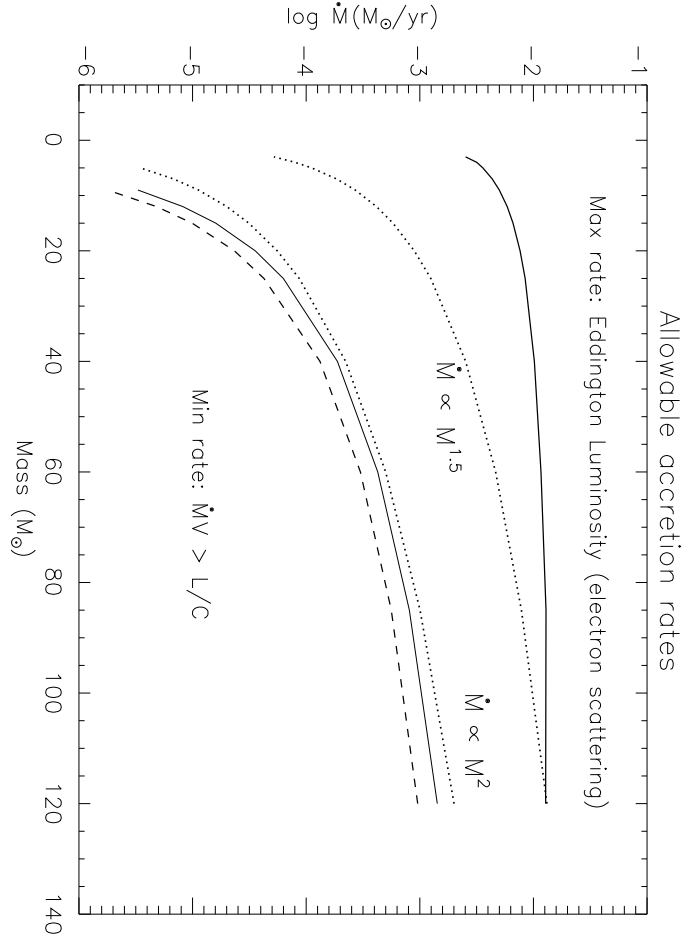}
\caption{
This figure shows the 
minimum and maximum accretion rates (solid lines) allowed by the
radiation of a high mass star and accretion shock.
The lower 
limit is set by the requirement that the inward momentum of the 
accretion flow exceed the outward momentum of the
radiation from the star and accretion shock. The upper limit on the
accretion rate is set by the requirement that the luminosity of the
star and shock be less than the Eddington luminosity of the star.
The numerical value for the cross-section for electron
scattering is taken from Lamers (1986).
An accretion rate suggested in this paper of $4.7\times 10^{-5}$
M$_\odot$ yr$^{-1}$ for an 18 M$_\odot$ star 
and scaling as M$^2$, and an accretion
rate suggested by Norberg and Maeder (2000),
$10^{-5}$ M$_\odot$ yr$^{-1}$ for a 1 M$_\odot$ star
and scaling as M$^{1.5}$ both (dotted lines) fall within the 
allowed region. Also plotted for reference is 
the minimum accretion rate suggested
by Wolfire and Cassinelli (1987) (dashed line). 
}

\end{figure}

\begin{figure}
\epsscale{0.6}
\plotone{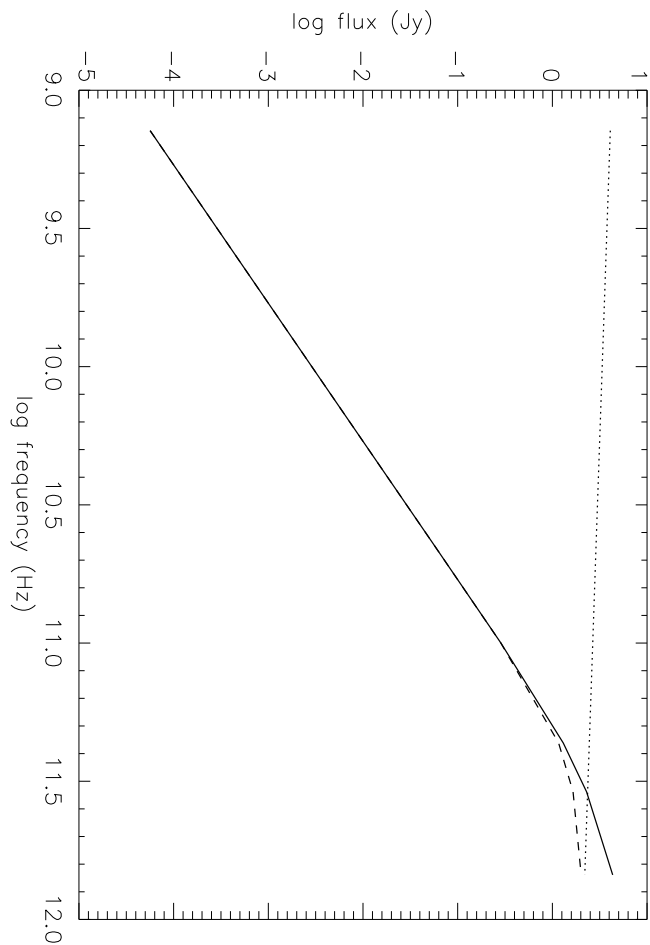}
\caption{
Predicted spectral index of a hypercompact HII region with radius of 107 AU
and an ionized density at the boundary of $2.5\times 10^7$ cm$^{-3}$. This
ionized gas density would be consistent with a 
molecular gas density at the boundary of $6.4\times 10^5$ cm$^{-3}$
and a molecular density of $2.1 \times 10^4$ cm$^{-3}$ at 0.1 pc.
The solid line shows the spectral index for an HII region with an $r^{-3/2}$
density gradient and the dashed line for an HII region with constant density.
Both HII regions are optically thick at all but the highest frequencies.
The dotted line shows the predicted flux using the optically thin approximation.
}

\end{figure}

\begin{figure}
\epsscale{0.6}
\plotone{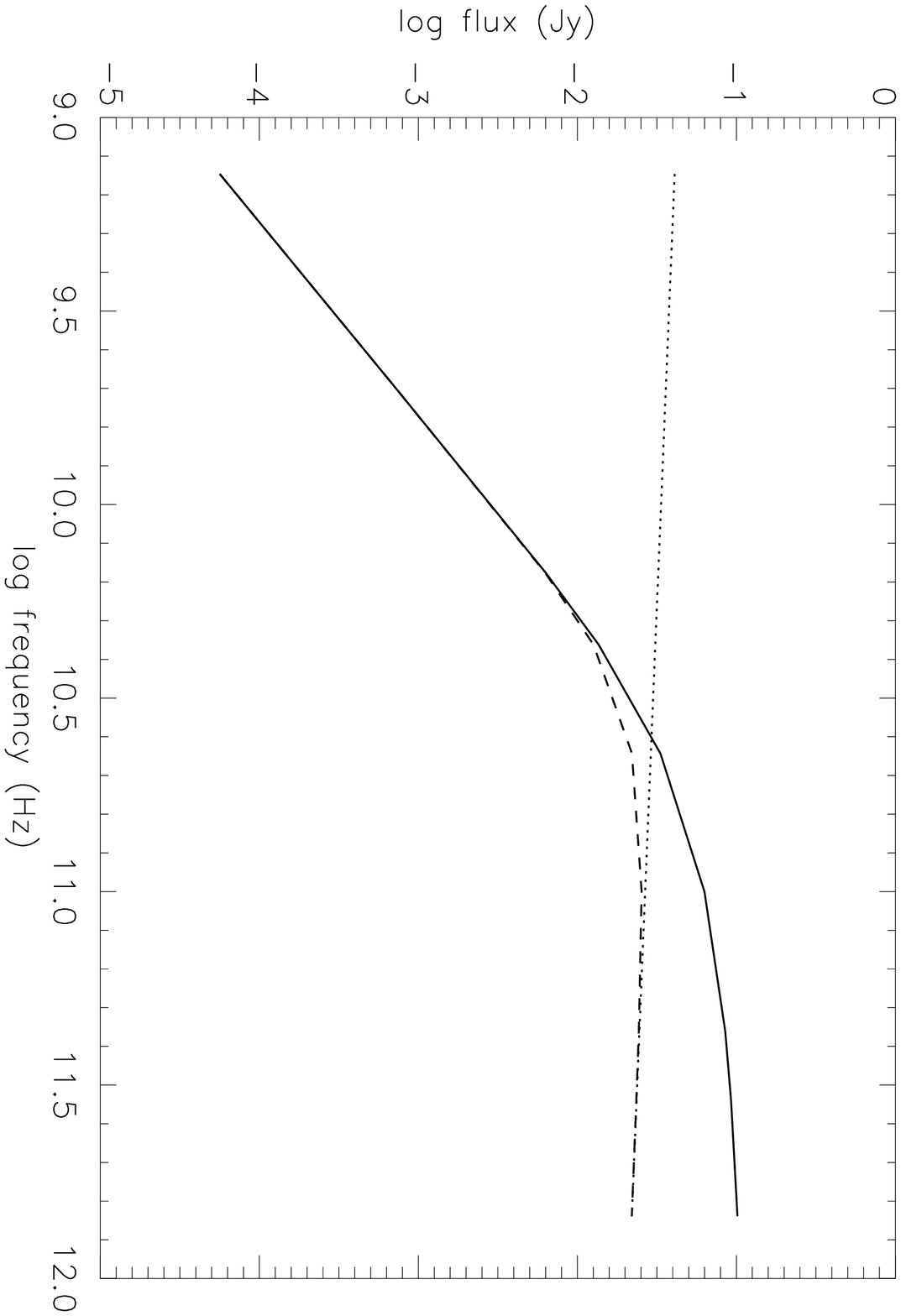}
\caption{
Predicted spectral index of a hypercompact HII region with radius of 107 AU
and a ionized density at the boundary of $2.5\times 10^6$ cm$^{-3}$ in the
same format as figure 3. 
}

\end{figure}

\begin{figure}
\epsscale{0.6}
\plotone{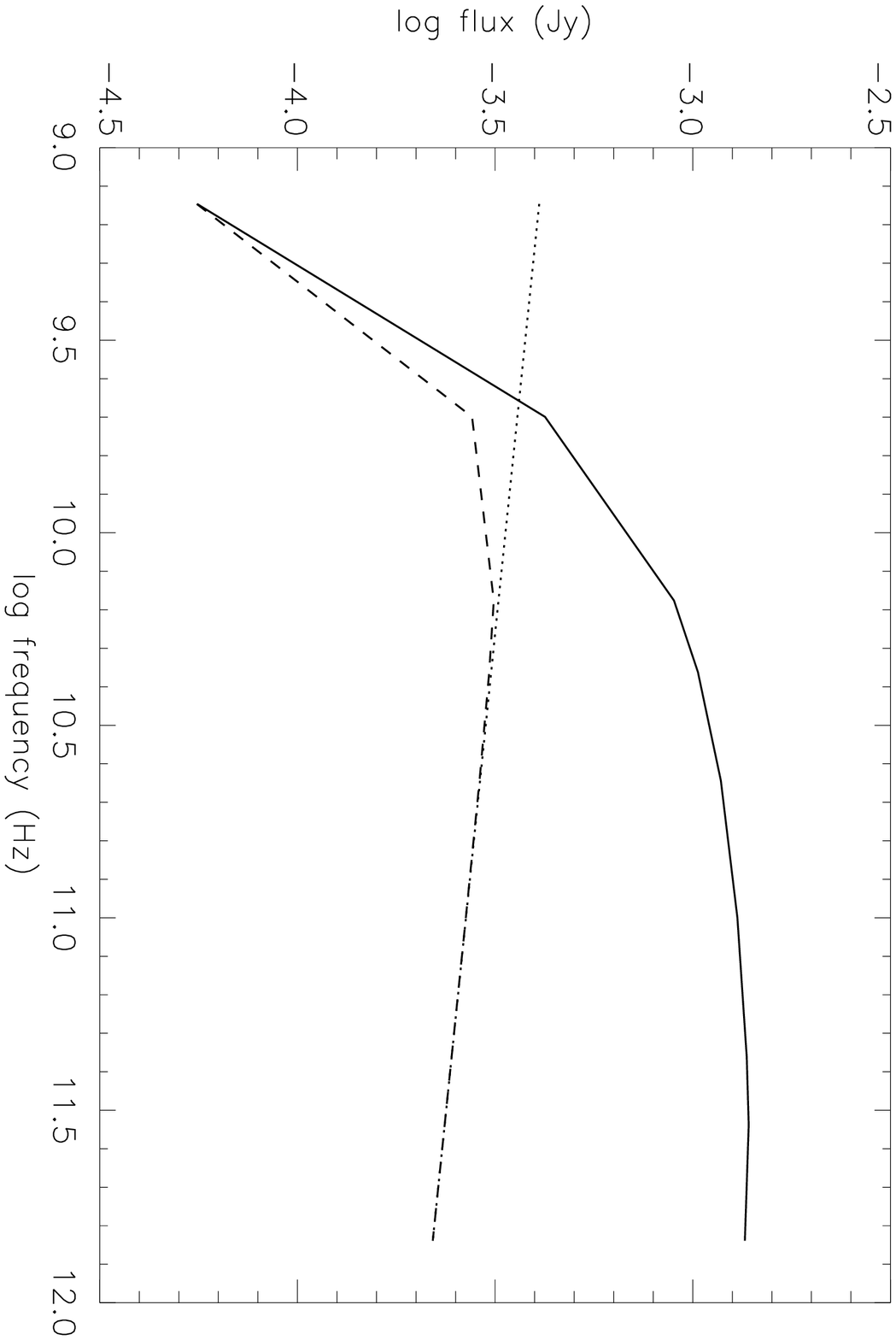}
\caption{
Predicted spectral index of a hypercompact HII region with radius of 107 AU
and an ionized density at the boundary of $2.5\times 10^5$ cm$^{-3}$ in the same
format as figure 3. 
}

\end{figure}


\clearpage

\begin{references}
\def\apj#1,    {{Ap.~J.{\rm,}\ {#1}{\rm,}\ }}
\def\apjl#1,   {{Ap.~J.~(Letters){\rm,}\ {#1}{\rm,}\ }}
\def\apjs   {{Ap.~J.~Suppl.{\rm,}\ }}
\def\aj#1,     {{A.~J.{\rm,}\ {#1}{\rm,}\ }}
\def\aa#1,     {{Astr.~Ap.{\rm,}\ {#1}{\rm,}\ }}
\def\pasj#1,     {{Publ.~Astron.~Soc.~Japan{\rm,}\ {#1}{\rm,}\ }}
\def\pasp#1,     {{Publ.~Astron.~Soc.~Pacific{\rm,}\ {#1}{\rm,}\ }}
\def\aas#1,    {{Astr.~Ap.~Suppl.{\rm,}\  {#1}{\rm,}\ }}
\def\araa#1,  {{Ann.~Rev.~Astr.~Ap.{\rm,}\ {#1}{\rm,}\ }}
\def\baas   {{Bull.~AAS{\rm,}\ }}
\def\mnras#1,  {{M.~N.~R.~A.~S.{\rm,}\ {#1}{\rm,}\ }}
\def\nature    {{Nature{\rm,}\ }}
\def\science#1,    {{Science{\rm,}\ {#1}{\rm,}\ }}
\def\vol#1  {{{#1}{\rm,}\ }}
\def\ref#1  {\noindent \hangindent=24.0pt \hangafter=1 {#1} \par}
\ref{Altenhoff, W., Strittmatter, P., \& Wendker, H., 1981, \aa 93, 48}
\ref{Appenzeller, I., \& Tscharnuter, W., 1974, \aa 30, 423}
\ref{Beech, M., \& Mitalas, R., 1994, \apjs 95, 517}
\ref{Behrend, A. \& Maeder, A., 2001, \aa 373, 190}
\ref{Bernasconi, P. \& Maeder, A., 1996, \aa 307, 839}
\ref{Bondi, M., 1952, \mnras 112, 195}
\ref{Carral, P., Kurtz, S., Rodriguez, L., Marti, J., Lizano, S., Osorio, M., 
1999, RMxAA, 35, 97}
\ref{Chieffi, A., Straniero, O. \& Salaris, M., 1995, \apjl 445, 39}
\ref{D'Antona, F. \& Mazzitelli, I., 1996, \apjs 90, 467}
\ref{dePree, C., Goss, W., Palmer, P., \& Rubin, R., \apj 428, 670}
\ref{dePree, C., Gaume, R., Goss, W., \& Claussen, M., 1996, \apj 464, 788}
\ref{dePree, C., Mehringer, D., \& Goss, W., 1997, \apj 482, 307}
\ref{Dyson, J. \& Williams, D., 1980, The Physics of the
Interstellar Medium (New York:Wiley)}
\ref{Franco, J., Tenorio-Tagle, G. and Bodenheimer, P. 1990, \apj, 349, 126}
\ref{Garay, G., Rodriguez, L., Moran, J., \& Churchwell, E., 1993, \apj 418, 368}
\ref{Garay, G., \& Lizano, S., 1999, \pasp 111, 1049}
\ref{Gaume, R., Fey, A., \& Claussen, M., 1994, \apj 233, 115}
\ref{Gaume, R., Claussen, M., dePree, C., Goss, W., \& Mehringer, D., 1995, \apj 449, 663}
\ref{Jaffe, D., Martin-Pintado, J., 1999, \apj 520, 162}
\ref{Jijina, J. \& Adams, F., 1996, \apj 462, 874}
\ref{Kahn, F., 1974, \aa 37, 149}
\ref{Ho, P., Klein, R., \& Haschick, A., 1986, \apj 305, 714}
\ref{Keto, E., Welch, W., Reid, M., Ho, P., 1995, \apj 444, 765}
\ref{Keto, E., 2002a, ApJ, 568, 754 (K1)}
\ref{Keto, E., 2002b, \apj 580, 980 (K2)}
\ref{Krugel, E. \& Siebenmorgan, R., 1994, \aa 288, 929}
\ref{Kurtz, S., Churchwell, E., \& Wood, D., 1994, \apjs 91, 659}
\ref{Lamers, H., 1986, \aa 159, 90}
\ref{Larson, R., \& Starrfield, S., 1971 \aa 13, 190}
\ref{Limongi, M., Straniero, O. \& Chieffi, A., 2000, \apjs 129, 625}
\ref{Lizano, S., Galli, D., Shu, F. \& Canto, J., 2003, RevMexAA, 15, 166}
\ref{Mathis, J., Rumpl, W., \& Nordsieck, K., 1977, \apj 217, 425}
\ref{Mestel, L., 1954, \mnras 114, 437}
\ref{Meynet, G., \& Maeder, A., 2000, \aa 361, 101}
\ref{Mezger, P., \& Henderson, A., 1967, \apj 147, 471}
\ref{Miralles, M., Rodriguez, L., \& Scalise, E., 1994, \apjs 92, 173}
\ref{Molinari, S., Brand, J., Cesaroni, R., Palla, F., Palumbo, G.,
1998, AA, 336, 339}
\ref{Nakano, T., 1989, \apj 345, 464}
\ref{Nakano, T., Hasagawa, T., Norman, C., 1995, \apj 450, 183}
\ref{Norberg, P. \& Maeder, A., 2000, \aa 359, 1025}
\ref{Panagia, N., 1973, \aj 78, 929}
\ref{De Pree, C., Goss, W., Palmer, P., \& Rubin, R., 1994, \apj 283, 632}
\ref{De Pree, C., Gaume, R., Goss, W., \& Claussen, M., 1996, \apj 464, 788}
\ref{Schaller, G., Schaerer, D., Meynet, G., Maeder, A., 1992, \aas 96, 269}
\ref{Shu, F., 1992, The Physics of Astrophysics, Volume II,
Gas Dynamics, University Science Books, Mill Valley, CA}
\ref{Shu, F., Lizano, S., Galli, D., Canto, J. \& Laughlin, G., 2002, \apj 580, 969}
\ref{Spitzer, L., Jr., 1978, Physical Processes in the Interstellar
Medium, (New York:Wiley)}
\ref{Tenorio-Tagle, G., 1979, \aa 71, 59}
\ref{Vacca, W., Garmany, C., Shull, J., 1996, \apj 460, 914}
\ref{Walmsley, M., 1995, Revista Mexicana de Astronomia y Astrofisica Serie de Conferencias, Vol. 1, Circumstellar Disks, Outflows and Star Formation,
Cozumel, Mexico, Nov 28-Dec 2, 1994, p. 137}
\ref{Wolfire, M. \& Cassinelli., 1986, \apj 310, 207}
\ref{Wolfire, M. \& Cassinelli., 1987, \apj 319, 850}
\ref{Wolfire, M. \& Churchwell, E., 1994, \apj 427, 889}
\ref{Wood, D. \& Churchwell, E., 1989a, \apjs 69, 831}
\ref{Wood, D. \& Churchwell, E., 1989b, \apj 340, 265}
\ref{Yorke, H., 1984, Workshop on Star Formation, ed R.D. Wolstencroft
(Edinburgh: Royal Obs.), 63}
\ref{Yorke, H., \& Krugel, E., 1977, \aa 54, 183}
\ref{Yorke, H., 2001, Hot Star Workshop III: The Earliest Stages
of Massive Star Birth, A.S.P. Conf. Ser. 267, 165}
\ref{Yorke, H., and Sonnhalter, C., 2002, \apj 569, 846}
\ref{Zijlstra, A., Pottasch, S., Engels, D., Roelfsma, P., te Lintel Hekkert, P.,
Umana, G., 1990, \mnras 246, 217}



\end{references}
\end{document}